# Water–Energy–Data nexus reveals sustainability gaps and priorities for AI infrastructure

Jincheng Huang[1, *], Guoan Xiao[1], Xin Lan[1], Caichun Yin[2, 3], Zhimeng Jiang[4], Yue Cui[5], Jianguo Liu[1, *]

1. Center for Systems Integration and Sustainability, Michigan State University, East Lansing, MI, USA

2. State Key Laboratory of Earth Surface Processes and Disaster Risk Reduction, Faculty of Geographical Science, Beijing Normal University, Beijing, China.

3. Institute of Land Surface System and Sustainable Development, Faculty of Geographical Science, Beijing Normal University, Beijing, China.

4. Department of Urban Planning and Design, The University of Hong Kong, Hong Kong SAR

5. School of Planning, Design, and Construction, Michigan State University, East Lansing, MI, USA.

*Correspondence: huang196@msu.edu (J.H.); liuji@msu.edu (J.L.)

## Abstract:

The rapid growth of artificial intelligence (AI) depends on interconnected water, energy, and data systems spanning infrastructure life cycles and diverse geographies, with the potential to intensify water and energy insecurity, exacerbate social inequities and lead to cascading sustainability issues globally. Yet assessments commonly examine them separately, limiting our understanding of where resource pressures arise, how they are transferred across space, and who bears their consequences. To address these challenges, we established a water–energy–data (WED) nexus framework and use large language models to synthesize 12, 825 water- or energy-related studies across mineral extraction, semiconductor manufacturing, data center operation and electronic-waste processing. The evidence map identifies a mismatch between the interconnected activities supporting AI infrastructure and boundaries used to assess them. Researchers assess operational energy more than water use and other indirect demands, while interactions across locations and life-cycle activities remain understudied. These gaps obscure how resource demands shift among infrastructure nodes and regions, and how data-enabled benefits compare with their underlying costs. The WED nexus provides a foundation for integrated assessment of spatially metacoupled systems, while the evidence map can inform priorities for managing the sustainability implications of the AI boom

## Introduction

Artificial intelligence (AI) is increasingly deployed to support sustainable development globally, including the management of water and energy systems[1,2]. Its rapid growth, however, relies on physical infrastructure that generates significant environmental and social pressures[3,4]. For instance, data centers consumed approximately 415 TWh globally (1.5% of total demand) in 2024[5], with U.S. facilities alone accounting for 192 TWh (4.7% of national electricity) projected to surge to 649 TWh (11.8%) by 2030[6]. Additionally, U.S. data centers consumed roughly 66 billion litres of water directly and an estimated 12 times that amount indirectly in 2023[7]. While existing assessments provide important evidence on such facility-level electricity demand and cooling water use[8–10], these metrics alone cannot fully capture AI's sustainability consequences. Reductions in direct facility cooling water can be offset by increased electricity demand and water consumption at power plants[11]. Beyond these systemic trade-offs, local vulnerabilities are exacerbated because data center capacity is highly clustered[12], with two-thirds of post-2022 facilities located in water-stressed regions[13]. Furthermore, AI services depend on the broader digital supply chain, such as semiconductor manufacturing, mineral extraction and the management of discarded equipment, connecting demand for computing to resource use and waste burdens across multiple regions[14–17]. Therefore, understanding how AI affects sustainability requires a comprehensive examination of the lifecycle of resource systems and local communities that support its infrastructure, weighed alongside the potential benefits of AI applications.

To address this complexity, we conceptualize the Water–Energy–Data (WED) nexus as the dependencies, interactions and feedbacks among water systems, energy systems, and data systems across the activities that support AI services (Fig. 1a). Building on the water–energy nexus, which explains the water requirements for energy production and energy requirements for water provision[18], the WED nexus extends it by introducing data and computing infrastructure as a third interacting system. These components connect demand for AI services to resource consumption while supporting monitoring and management of water and energy systems[19,20]. Treating data infrastructure as part of the nexus therefore expands the analytical boundary of the issue: it connects resource consumption at computing facilities with upstream production, downstream waste, regional resource constraints and the potential feedback through which water and energy systems may themselves limit future infrastructure development.

While the WED nexus captures the systemic dependencies within specific places, we organize these relationships across space (within and across different places) using the metacoupling framework[21]. The metacoupling framework captures how human–environment interactions are distributed within systems (e.g., places), between adjacent systems, and across distant systems, making it particularly suited to analyze AI infrastructure whose resource demands and environmental consequences are spatially decoupled from end-users[3]. A data center provides an illustrative focal system—cooling and computing occur within the facility, utilities and communities connect it to surrounding systems, and hardware production and end-of-life activities can connect it to distant locations (Fig. 1b). Extended definitions and illustrative pathways are provided in Supplementary Note 1 and Supplementary Table 1.

Here, we conduct a large language model (LLM)-assisted structured synthesis of 12,825 unique water- or energy-related studies across four resource-intensive AI infrastructure activities: mineral extraction, semiconductor manufacturing, data center operation and e-waste processing. We quantify gaps in the evidence base by examining how the retrieved quantitative literature reports direct and indirect resource use, water–energy relationships, pressures on nearby resource systems, connections across infrastructure activities, and environmental and social impacts and feedbacks. The analysis maps differences in coverage across resource domains, life-cycle activities, and spatial scales. We interpret these patterns through the WED nexus to identify priorities for integrated assessment and inform governance of AI's implications for sustainability.

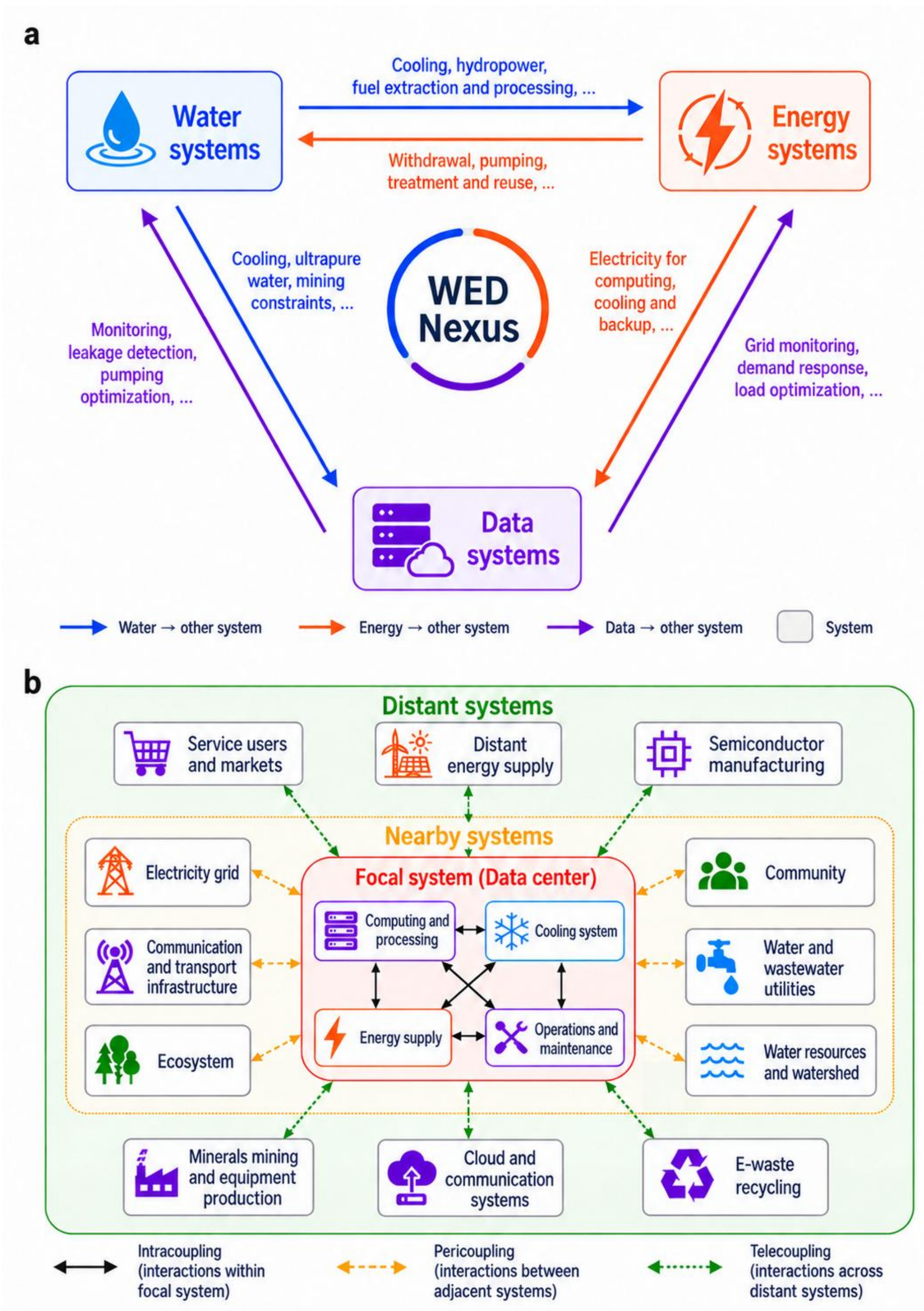

**Figure 1. Conceptual framework of the WED nexus. a**, the bidirectional pathways show that each system both enables and constrains the other two. Water systems support energy production and data infrastructure through cooling, hydropower, fuel processing and water supply; energy systems power water withdrawal and treatment as well as computing, cooling and backup systems; and data systems support monitoring, forecasting and optimization in both water and energy systems. Blue, orange and purple arrows indicate

flows originating from water, energy, and data systems, respectively. **b**, A data center is used as an illustrative focal system to show how WED interactions can be organized across spatial scales. Intracoupling captures interactions within the focal system, pericoupling links the data center with adjacent water, energy, infrastructure, community and ecosystem systems, and telecoupling connects it to distant supply chains, energy sources, digital networks, markets and end-of-life processes.

# Results

## *Rapid growth and geographical unevenness of the evidence base*

To quantify gaps in the evidence base, a structured coding framework was conducted to evaluate how existing research addresses the WED nexus across four resource-intensive AI infrastructure nodes: mineral extraction, semiconductor manufacturing, data centers, and e-waste processing. Publication output across the water-data (WD), energy-data (ED), and overlapping WED corpora rose overall, with year-to-year fluctuations and a marked acceleration after 2023 (Fig. 2a–c) aligning with the generative AI boom. However, the scale and composition of the evidence remain highly uneven. Energy-related research constitutes by far the largest share of the evidence base and is dominated by data center studies, reflecting a strong focus on electricity consumption, operational efficiency and cooling demand in assessments of AI infrastructure. Water-related research is less and more distributed across the life cycle, with mineral extraction contributing to a substantial share of recent growth. Studies that jointly consider water and energy remain comparatively limited, despite their rapid expansion after 2023.

The distribution of research across the four nodes also reveals a fragmented understanding of resource demand. Data centers dominate energy-related studies, whereas upstream mineral extraction accounts for a much larger share of water-related and integrated water–energy research. Semiconductor manufacturing remains weakly represented across all three corpora, despite the intensive electricity and ultrapure water requirements of semiconductor fabrication. E-waste processing contributes to a growing but still comparatively small body of evidence. Research attention therefore does not follow the AI infrastructure life cycle evenly: different resource pressures are examined in different node-specific literatures, rather than as connected demands generated by a common infrastructure system.

Geographical coverage is similarly uneven (Fig. 2d, e). Among the 12,825 unique eligible studies, 1,863 (14.53%) reported identifiable country information in their titles or abstracts. Country-level studies are concentrated in the United States, China, Europe and a limited number of countries in East and South Asia. The maps compare these counts with national Water Resource Stability and Energy Trilemma scores[22,23]. Some countries with lower indicator scores had few geographically identifiable studies, highlighting locations for further evidence of retrieval. These patterns show that the evidence base is expanding rapidly without becoming proportionately more balanced. Research remains concentrated on data center energy use, selected upstream water pressures, and a relatively small group of countries. As a result, current scholarship provides an incomplete basis for comparing resource dependence across the principal water- and energy-intensive nodes of AI infrastructure or for evaluating how their expansion interacts with heterogeneous resource constraints worldwide.

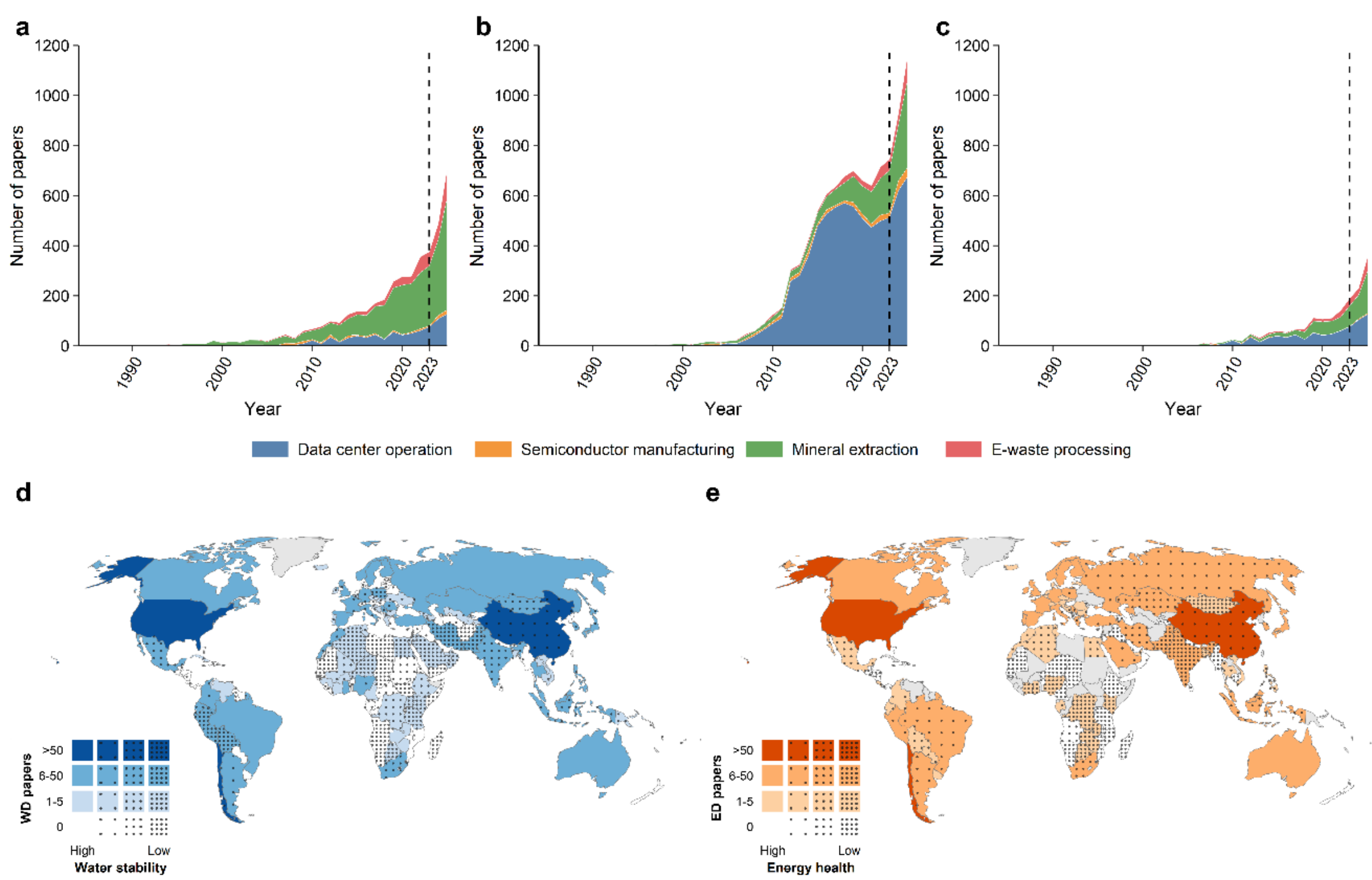

**Fig. 2 | Temporal and geographic patterns of resource-related research across AI infrastructure. a–c**, Annual publication counts for water–data (WD), energy–data (ED)

and water–energy–data (WED) studies, respectively, through December 2025. Stacked areas distinguish data center operation, semiconductor manufacturing, mineral extraction and e-waste processing; the dashed line marks 2023. **d**, National water-resource stability and WD research coverage. **e**, Energy System health and ED research coverage. Fill color indicates country-level study, whereas dot density indicates lower scores on the corresponding water or energy indicator.

### *Limited attention to indirect water and energy consumption*

Energy-related interactions are quantified much more consistently than water-related interactions, particularly at the operational stages of AI infrastructure (Fig. 3). At the data center node, 70.60% of ED papers quantify direct energy or electricity use, compared with only 24.53% of WD studies that quantify water withdrawal, water consumption, or cooling-water flow. Direct energy quantification is similarly prominent in chip-manufacturing studies (73.7%) and remains substantial in the mining (54.2%) and e-waste (52.8%) literature. This pattern is consistent with the treatment of electricity demand and energy efficiency as core performance variables across AI infrastructure, whereas water use remains more unevenly measured and is often framed as a secondary requirement of cooling or production processes.

The analytical boundary varies among infrastructure nodes. In energy-focused data center and semiconductor studies, 11.99% and 6.58%, respectively, quantify energy use for water systems, while 0.75% and 3.84% quantify other life-cycle energy requirements. Other life-cycle energy use is reported more often in mineral extraction and e-waste processing, at 14.92% and 21.97%, respectively. The water-focused data center literature remains particularly concentrated on direct operations: 21 of 856 studies (2.45%) quantify water use for energy, and five (0.58%) quantify other life-cycle water use. Despite often being magnitudes larger than direct use[7], indirect water use remains disproportionately overlooked. These results reveal not only an imbalance between energy and water quantification, but also a fragmented division of analytical attention across the AI-infrastructure life cycle: operational energy use dominates research on data centers and semiconductor manufacturing, whereas upstream and downstream studies more often address life-cycle burdens within their own nodes.

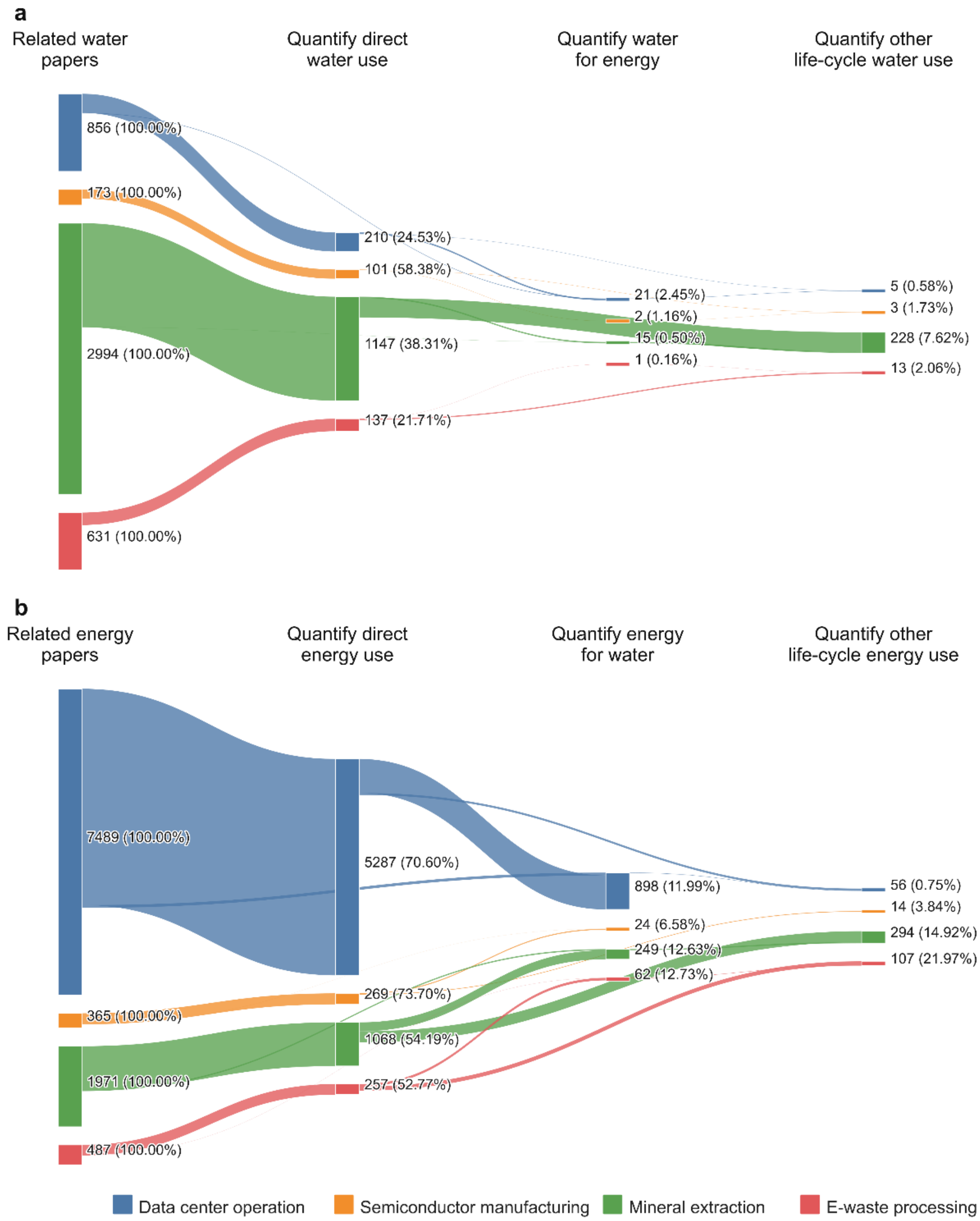

**Fig. 3 | Limited attention to indirect water and energy consumption in WED nexus studies.**

**a,** Share of water-related AI-infrastructure studies that quantify direct water use, water use for energy, and other life-cycle water use. **b,** Share of energy-related AI-infrastructure studies that quantify direct energy use, energy use for water, and other life-cycle energy use. The flows indicate the number and percentage of studies in each category, showing that existing research is dominated by direct use accounting, whereas indirect water and energy consumption associated with energy production, water-cooling systems, and broader life-cycle processes remains rarely quantified.

### *Little assessment of nearby and distant resource pressures*

Assessment of resource use remained concentrated within facility boundaries, with limited attention to pressures on surrounding water and energy systems (Fig. 4a). Facility-level assessment covered 21.71-58.38% of WD studies and 52.77-73.70% of ED studies across the four AI infrastructure nodes. By comparison, nearby water pressures were assessed in only 7.01% of mineral extraction studies, 6.18% of e-waste processing studies, 4.62% of semiconductor-manufacturing studies, and 1.75% of data center studies. Nearby energy pressures received even less attention, appearing in 1.88% of mineral extraction studies and 1.38% of data center studies, and in none of the eligible semiconductor-manufacturing or e-waste processing studies. Resource demand was therefore assessed much more frequently within individual facilities or production processes than in relation to the availability, capacity, or competing demands of surrounding resource systems.

Coverage extending to other infrastructure nodes or broader life-cycle activities was also uneven and did not consistently decline with increasing analytical scope. Such coverage reached 13.43% in WD mineral extraction studies, compared with 4.12% for e-waste processing, 3.47% for semiconductor manufacturing, and 0.70% for data centers. In the ED literature, it reached 28.13% for e-waste processing and 23.14% for mineral extraction, but only 8.49% for semiconductor manufacturing and 0.87% for data centers. These comparatively high values in the mining and e-waste literature exceeded their coverage of nearby resource pressures, indicating that an expanded life-cycle boundary does not necessarily entail greater attention to host region resource constraints.

The directional matrix further reveals that cross-node assessment was concentrated on a small number of connections (Fig. 4b). The most frequently examined connection ran from mineral extraction to e-waste processing, appearing in 71 WD studies (2.37%) and 149 ED studies (7.56%). ED semiconductor studies most often connect manufacturing energy use to data center operation, whereas ED e-waste studies most often connect processing energy use to mineral production. By contrast, no WD data center study quantified a connection with any of the other three nodes, and ED data center connections with semiconductor manufacturing and mineral extraction appeared in only 0.05% and 0.04% of studies, respectively. Current research therefore captures selected material recovery and production relationships without providing a balanced account of resource flows across extraction, semiconductor fabrication, data center operation, and end-of-life processing. Together, the limited assessment of host region pressures and the selective treatment of cross-node connections leave substantial uncertainty about how water and energy burdens accumulate, interact and shift between regions across the AI infrastructure life cycle.

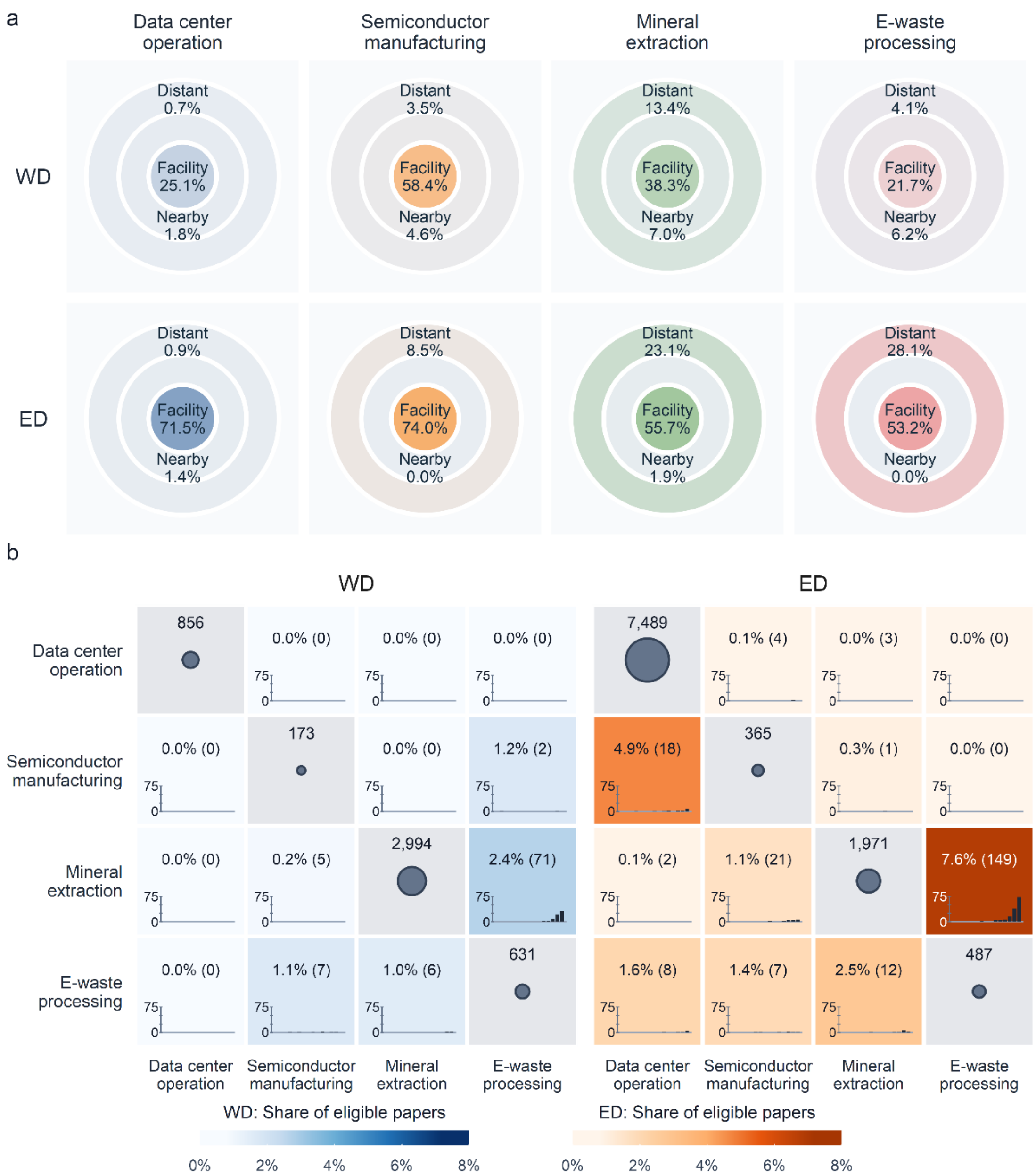

**Fig. 4 | Little and uneven assessment of nearby and distant resource pressures. a**, for each focal-node literature, concentric rings show the share of eligible papers assessing resource consumption within facilities (inner), in nearby systems (middle), and across distant focal nodes (outer). Node hue identifies the focal system, and color intensity denotes prevalence on a shared 0-80% scale. **b**, Directed matrix of cross node assessments. Rows denote the focal node literature and columns from the external focal node assessed; color and labels show the percentage and number of eligible papers quantifying each linkage. Diagonal circles are scaled by the total number of eligible papers for each focal node of literature. Black mini bars show the number of papers coded as quantifying a directed linkage within consecutive three-year periods from 1981 to 2026; their vertical scale is shared across all cells and panels.

### *Gaps in environmental and social impacts and feedback*

Both environmental and social interactions were seldom assessed across the four types of AI infrastructure, with social interactions receiving substantially less attention than environmental interactions (Fig. 5). Environmental interactions were assessed in 7.81% of WD studies and 12.00% of ED studies, compared with only 1.05% and 1.58%, respectively, for social interactions. This difference was present at every kind of infrastructure. Environmental interactions were most frequently assessed in water use of mineral extraction and energy use of e-waste processing studies, but even these relatively high values represented only a minority of the relevant literature. Overall, environmental interactions received approximately 7.5 times more attention than social interactions, indicating that the evidence base privileges biophysical processes over effects on communities, livelihoods, institutions and resource access.

Interactions between distant systems were even less frequently studied than those between adjacent systems. Across the four nodes, nearby interactions appeared in 7.81% of WD studies and 11.97% of ED studies, whereas distant interactions appeared in only 0.39% and 2.26%, respectively. E-waste was the only node for which distant ED environmental-impact coverage slightly exceeded nearby coverage, reflecting comparatively greater attention to environmental conditions and consequences across end-of-life supply chains. This broader spatial reach did not extend to social interactions. Only two ED mineral extraction studies assessed distant social impacts, and none assessed distant social feedback. Research extending beyond host regions therefore remains concentrated in selected environmental dimensions of resource intensive supply chains, while distant communities and ecosystems are almost entirely absent.

Feedbacks from environmental or social systems were less frequently studied than outward impacts. Impacts of energy use from e-waste processing were assessed in 20.53% of studies but feedbacks in only 0.62%, and in water use of mineral extraction studies, where the corresponding values were 9.49% and 0.67%. Data center research differed from this general pattern. In the WD literature, feedback coverage reached 3.04%, exceeding impact coverage of 1.75%, while ED data center studies showed nearly equal coverage of impacts and feedbacks at 7.24% and 7.10%. This relative attention to feedback may reflect the operational sensitivity of data centers to climate, water availability, electricity systems, and utility conditions. Nevertheless, studies rarely integrated both directions: the highest bidirectional coverage was only 0.96% in ED data center research, and no study assessed bidirectional social interactions. These gaps limit the ability of current research to explain how resource pressures translate into unequal exposure, costs and benefits, and how affected communities and institutions subsequently influence infrastructure development and resource demand.

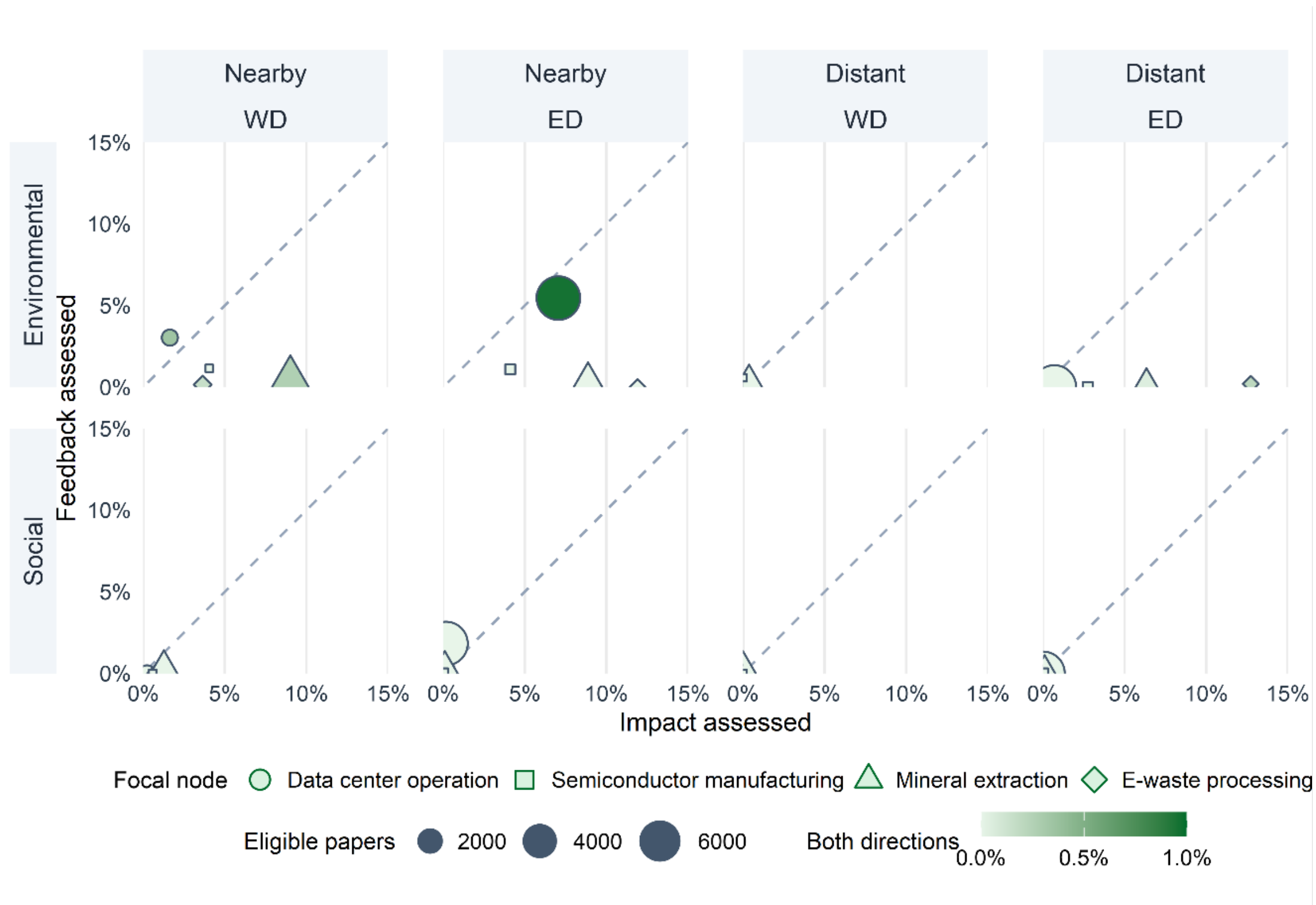

**Fig. 5 | Uneven assessment of environmental and social impacts across the AI-infrastructure life cycle.** Each point represents the literature associated with one focal node in a WD or ED literature set. The x- and y-axes show the shares of eligible papers assessing resource impacts and feedbacks, respectively; the dashed line indicates equal directional coverage. Point shape identifies the focal node, point area represents the number of eligible papers, and green fill denotes the share of papers assessing both directions. Columns distinguish nearby from distant interactions and WD from ED papers, whereas rows distinguish environmentally from social interactions.

# Discussion

## *Fragmented evidence establishes the need for the WED nexus*

The evidence map identifies a mismatch between the interconnected activities supporting AI infrastructure and the boundaries used to assess them. Direct energy use is prominently represented in data center and semiconductor research, while broader life-cycle requirements receive greater attention in parts of the mining and e-waste literature. Yet broader life-cycle coverage does not consistently include pressures on nearby resource systems. This distinction matters because accounting for an upstream or downstream process and evaluating the capacity of the system that hosts it answers different questions. The findings point to a need to connect life-cycle accounting with assessment of resource availability, competing demands and feedbacks.

This fragmentation matters first because the infrastructure nodes in the lifecycle of AI services are materially interdependent. Computing demand stimulates hardware production[6,17]; semiconductor fabrication requires electricity, chemicals, minerals and ultrapure water[24]; equipment replacement creates downstream recovery and waste-processing requirements[17]; and electricity generation connects data center operation to water consumption beyond facility boundaries[9]. Furthermore, such fragmentation obscures the social and environmental consequences of distant interactions. Digital services are not immaterial flows but outputs of resource-intensive infrastructures. Data and reusable digital knowledge can exhibit quasi-public-good characteristics with broadly distributed benefits[25], while their supporting infrastructure imposes resource pressures and environmental burdens on particular production and host communities[26,27]. Efficiency gains can lower service costs and stimulate further demand—the mechanism underlying the Jevons paradox—so lower resource use per service does not necessarily translate into lower total use[28]. When growing infrastructure demand places pressure on shared resources without accounting for costs borne elsewhere, weak coordination may create common-pool-resource conflicts and, under sufficiently weak coordination, tragedy-of-the-commons dynamics[29]. Facility-level evidence alone cannot reveal these aggregate effects or clarify how responsibility should be shared among the actors that support and use AI services[30].

The WED nexus responds to this fragmentation by treating AI services and their supporting infrastructure as a network of connected resource flows. It connects facility operations to wider water–energy dependencies, including water consumed in electricity generation and energy required for ultrapure-water production. It follows these demands through the infrastructure life cycle, from upstream production to equipment recovery[14,17,31,32]. This nexus framework makes cross-sectoral and cross-regional transfers visible. Additionally, it allows findings from different nodes to be interpreted in their local and temporal contexts. For instance, ore grade and hydrology[33,34], fabrication processes and recycling rates[24,31], workloads and cooling strategies[4,9], and waste treatment pathways all change resource demand[17]. AI workloads and hardware replacement patterns further shape these demands across the infrastructure life cycle[14,17]. The framework thus brings otherwise separate estimates into an account of where, when, and through which processes water and energy pressures arise, including during drought, heat waves, and peak grid demand[9].

Beyond connecting infrastructure burdens, the WED nexus has the potential to capture the active role of data and computing in water and energy systems[19,35]. AI-enabled monitoring, forecasting and control can help electricity systems balance supply and demand and integrate renewable generation[20,35], while waste heat from data centers can provide a useful thermal-energy source[36]. Considering these benefits alongside the resources required to produce them enables assessment of net system effects rather than treating AI either solely as a burden or automatically as a solution[20,37].

## *Priorities for the WED nexus assessment and governance*

The WED nexus offers a way to assess AI services and infrastructure as coupled systems rather than separate demands for water, electricity, and computing. Similar to the Water-Energy-Food nexus, water and energy are not only essential to each other, but also both essential to the data system[38]. Future research should therefore trace how choices across the AI infrastructure life cycle alter several resource demands at once. Evaporative cooling, for example, can lower data center electricity demand while increasing direct water consumption and dependent on local water availability[9,36]. Dry or hybrid cooling can reduce direct water dependence, but it can increase electricity demand during hot periods and shift part of the burden from the water system to the power system[9]. Similarly, greater material recovery from mining or electronic waste can reduce resource losses while requiring additional water and energy[39,40]. These trade-offs are shaped by regional resource conditions, while their social consequences depend on how affected communities are exposed to the resulting pressures[4,40]. Assessments should examine marginal conditions alongside annual averages and evaluate net impacts across multiple objectives, including whether growing AI demand competes for electricity and infrastructure capacity needed for decarbonization[5].

A second priority is to trace how these trade-offs propagate across space through coupled infrastructure and supply-chain networks. Infrastructure expansion can reshape host-region water and energy systems, with consequences for the local environment and affected communities[3,9]. Those conditions can, in turn, influence infrastructure operations and workload management[14]. AI infrastructure also relies on distant systems for essential hardware, raw materials, and energy resources[14,16]. Consequently, a decision that improves performance at one facility may shift environmental footprints elsewhere[26]. This creates cascading effects that are not visible in facility-level assessments, requiring spatially explicit analysis of how system constraints move through metacoupled AI infrastructure networks.

This broader assessment should inform governance. Because social impacts remain insufficiently examined, WED assessments should address who bears burdens, whose contributions and concerns are recognized, and who can participate in decisions[42]. A flow-based approach can connect direct and indirect resource use, operational and embodied impacts, and local and transboundary burdens[43]. This requires coordination across the resource systems and supply chains supporting AI, informed by disclosure of their life-cycle resource demands.

Finally, WED research should test how these interconnected systems perform under concurrent disruptions. Concurrent shocks like climate change, geopolitical events, pandemics, and supply-chain shocks can disrupt the resource supplies and production networks on which AI infrastructure depends[44–46]. Heat waves and droughts jointly increase cooling demand while constraining water and electricity supply[27]; extreme weather damages power, transport, and communication infrastructure[47,48]; and pandemics or geopolitical disruptions delay semiconductor production, equipment supply, and system repair[49]. Because such disturbances can overlap and amplify one another[48], system resilience and sustainability depend not only on facility-level robustness, but also on the stability of interconnected water, energy, and data systems and their supply chains under concurrent stress.

*Limitations and Outlook*

This study was designed as a systematic evidence map to characterize the distribution and connectivity of a large literature rather than extract every result reported in every study. Systematic mapping commonly prioritizes consistent coverage of broad evidence bases and the identification of knowledge clusters and gaps[50]. Automated assistance, including machine learning and LLM-based approaches, has been developed to support efficient title-and-abstract screening in large evidence syntheses[51]. Accordingly, coding titles and abstracts offered a scalable and reproducible method for comparing our 12,825 records across multiple resource domains and infrastructure nodes. While human full-text validation confirmed that consensus among three LLMs achieves high accuracy, an inherent limitation is that assessments detailed exclusively within full texts may still be omitted. Regarding data sources, the Web of Science Core Collection was selected because it provides consistent bibliographic metadata across fields, although relevant grey literature[52], reports, and non-English publications may be underrepresented[53]. Similarly, the four selected nodes capture major resource-intensive activities associated with AI infrastructure but do not constitute an exhaustive life-cycle inventory (e.g., buildings, networking equipment, and transport may enter broader life-cycle categories without being searched as independent nodes[14]). Finally, because mineral extraction, semiconductor manufacturing, and e-waste processing also support non-AI activities, this synthesis maps evidence relevant to AI infrastructure but does not attribute all sectoral resources used exclusively to AI.

Future research can deepen this broad evidence map by conducting full-text syntheses utilizing more advanced LLMs and larger token allowances, and by incorporating additional databases that capture other infrastructures across the digital service life cycle. Studies distinguishing shared digital infrastructure from incremental requirements associated with specified AI workloads would further strengthen attribution[54].

Beyond these methodological refinements, our synthesis reveals a critical mismatch between the interconnected activities supporting AI and the fragmented ways researchers assess them. Despite rapid growth in literature following the AI boom, the focus on direct resource consumption of facility operations obscures how resource demands and their social and environmental interactions shift across host regions and distant supply chains. Furthermore, the evidence base exhibits a stark spatial imbalance, as research concentrates on major economies while resource-vulnerable regions—where many indirect impacts occur—remain underrepresented. The WED nexus brings these fragmented relationships into a unified frame. Looking forward, moving beyond isolated estimates to examine the complex interactions among AI infrastructure, global resource systems, and affected communities is essential to navigate the sustainability implications of the AI era.

53. 

## Acknowledgements

J.H. was supported by the Environmental Science and Policy Program and the Department of Fisheries and Wildlife at Michigan State University, as well as the National Science Foundation Long-term Ecological Research Program (DEB 2224712) at the Kellogg Biological Station. We are grateful for additional support from the National Science Foundation (grant 2118329), the USDA NIFA (2023-68012-39076), and Michigan AgBioResearch) (J.L.).

## Author information

### *Authors and Affiliations*

**Center for Systems Integration and Sustainability, Michigan State University, East Lansing, MI, USA**

Jincheng Huang, Guoan Xiao, Xin Lan, Jianguo Liu

**State Key Laboratory of Earth Surface Processes and Disaster Risk Reduction, Faculty of Geographical Science, Beijing Normal University, Beijing, China.**

Caichun Yin

**Institute of Land Surface System and Sustainable Development, Faculty of Geographical Science, Beijing Normal University, Beijing, China.**

Caichun Yin

**Department of Urban Planning and Design，The University of Hong Kong，Hong Kong SAR**

Zhimeng Jiang

**School of Planning, Design, and Construction, Michigan State University, East Lansing, MI, USA.**

Yue Cui